\documentstyle{mn}
\input{epsf}

\title[Phase-resolved spectroscopy of SAX J1808.4--3658]
{Phase-resolved X-ray spectroscopy of the millisecond pulsar SAX J1808.4--3658}

\author[M. Gierli\'nski, C. Done, D. Barret]
{Marek~Gierli\'nski$^{1, 2}$, Chris Done$^{1}$, Didier Barret$^{3}$\\
$^1$Department of Physics, University of Durham, South Road, Durham DH1 3LE, 
UK\\
$^2$Astronomical Observatory, Jagiellonian University, Orla 171, 30-244 
Krak{\'o}w, Poland\\  
$^3$CESR, 9 avenue du Colonel Roche, BP 4346-31028, Toulouse, Cedex
4, France}

\date{Submitted to MNRAS}
\pagerange{\pageref{firstpage}--\pageref{lastpage}}
\pubyear{2001}

\begin{document}

\topmargin = -0.5cm

\maketitle

\label{firstpage}

\begin{abstract}
 
  We present new results based on {\it RXTE\/} observations of the 
  millisecond pulsar SAX J1808.4--3658 carried out during the decay 
  of the April 1998 outburst. The X-ray spectrum can be fitted by a 
  two-component model. We interpret the soft component as blackbody 
  emission from a heated spot on the neutron star, and the hard 
  component as coming from Comptonization in plasma heated by the 
  accretion shock as the material collimated by the magnetic field 
  impacts onto the neutron star surface. The hotspot is probably the 
  source of seed photons for Comptonization.  The hard component 
  illuminates the disc, giving rise to a reflected spectrum. The 
  amount of reflection indicates that the disc is truncated at 
  fairly large radii (20--40 $R_g$), consistent with the lack of 
  relativistic smearing of the spectral features. The inferred 
  evolution of the inner radius is not consistent with the magnetic 
  field truncating the disc.  Instead it seems more likely that the 
  inner disc radius is set by some much longer time-scale process, 
  most probably connected to the overall evolution of the accretion 
  disc. This disc truncation mechanism would then have to be generic 
  in all low mass accretion rate flows both in disc accreting 
  neutron stars and black hole systems.

  The phase resolved spectra show clearly that the blackbody and hard
  Comptonized spectra pulse independently. This obviously gives an
  energy dependent phase lag. Full general relativistic effects are
  not required to explain this. The soft blackbody component is
  optically thick, so its variability is dominated by its changing
  projected area, while the Doppler shifts (which are maximized
  $90^\circ$ before the maximum in projected area) are somewhat
  stronger for the translucent column.

  We do not detect Compton reflection from the neutron star
  surface, though we predict that it should be present in the X-ray
  spectrum. This would give an unambiguous observational measure of
  $M/R$ if there is any iron on the neutron star surface which is not
  completely ionized. 
  
\end{abstract}


\section{Introduction}
\label{sec:introduction}

Neutron stars and Galactic black holes have very similar 
gravitational potentials, so the accretion flow in these systems 
might also be expected to be similar. The key difference is that 
neutron stars have a solid surface. Therefore, they can have a 
boundary layer and thermal emission from the surface.  Additionally, 
where there is a strong magnetic field the accretion flow is 
collimated along the field lines. When the spin and magnetic axes 
are misaligned the surface emission should be coherently pulsed at 
the spin period. This is seen in the X-ray pulsars -- high mass 
X-ray binary systems where young, rapidly rotating neutron stars 
with a high magnetic field ($\ge 10^{12}$ G) collimate the accreting 
material onto the magnetic poles. However, up till 1998 no {\em 
millisecond\/} pulses were seen from the low mass X-ray binaries 
(LMXBs), so there was no direct measure of the spin and magnetic 
field in such systems.  This would be an important test of models of 
the formation of the millisecond pulsars. Currently the most 
probable scenario for these rapidly rotating, low magnetic field 
($\sim 10^8$ G) neutron stars is that their progenitors are the 
LMXBs, where the neutron star is spun up through accretion torques, 
which also might dissipate the magnetic field (e.g. the review by 
Bhattacharya \& Srinivasan 1995).

This situation changed dramatically with the discovery of the coherent
X-ray pulsations with 2.5 ms period from the transient X-ray source
SAX J1808.4--3658 (Wijnands \& van der Klis 1998a). Its rapid rotation
and inferred magnetic field of a few $\sim 10^8$ G is in excellent
agreement with the predictions of the millisecond pulsar progenitor
models (Psaltis \& Chakrabarty 1999).  The key question is then what
makes SAX J1808.4--3658 so different from the rest of the LMXBs?
There are only a rather limited number of fundamental parameters which
determine the accretion flow -- the mass of the neutron star, mass
accretion rate, magnetic field, spin of the neutron star and angle
between the spin and magnetic axes. These could be coupled with the
age and/or evolutionary state of the system, and a further variable
which could alter the appearance of the system is its inclination
angle to the line of sight.

One way in which to look for differences is to do a detailed
comparison of the spectrum and variability of SAX J1808.4--3658 with
other neutron star LMXBs. These fall into two main categories, atolls
and Z sources, named after their different behaviours on a
colour--colour diagram (Hasinger \& van der Klis 1989). Atolls can
show rather hard spectra dominated by a power law component, similar
to that seen in the galactic black holes (e.g. Mitsuda et al.\ 1989;
Yoshida et al.\ 1993; Barret et al.\ 2000). This hard or island state
is generally seen only at mass accretion rates below a few per cent of
Eddington (Ford et al.\ 2000). Above this the source spectrum makes a
rapid transition to a much softer spectrum, termed the banana branch,
so tracing out a C (or atoll) shape on a colour--colour diagram.
Conversely the Z sources move from a horizontal branch, through a
normal branch to the flaring branch as their mass accretion rate
increases, tracing out a Z shape on a colour--colour diagram.  The
power spectral properties also correlate with atoll or Z source
classification, and with position on the colour--colour diagram (see
e.g. the review by van der Klis 1995).

The broad band spectrum of SAX J1808.4--3658 is entirely typical of an atoll 
system in the island state, with a power law hard X-ray spectrum (Heindl \& 
Smith 1998; Gilfanov et al.\ 1998). The variability power spectrum of SAX 
J1808.4--3658 is also very similar to island state atoll systems (apart from the 
coherent signal at the spin period): the broad band noise properties (Wijnands 
\& van der Klis 1998b) and even the correlation between the noise break 
frequency and the QPO frequency matches onto that from other LMXBs (Wijnands \& 
van der Klis 1999).

Thus it seems highly unlikely that there is any large scale difference between 
the properties of the accretion flow in SAX J1808.4--3658 compared with other 
neutron star LMXB systems. Yet there is a coherent spin pulse in SAX 
J1808.4--3658 which is definitely not present in the other atoll systems  
(Vaughan et al.\ 1994; Chandler \& Rutledge 2000). 

There were three observed outbursts of SAX J1808.4--3658: in September 1996 (in 
't Zand et al.\ 1998), in April 1998 (Wijnands \& van der Klis 1998a) and in 
January 2001 (Wijnands et al.\ 2001). Here we examine the {\it RXTE\/} spectra 
of SAX J1808.4--3658 from 1998 outburst in detail, looking at both the phase 
averaged and phase resolved spectra at differing mass accretion rates as the 
outburst declines. We derive a source geometry which is consistent with the 
observational constraints, and finally speculate that evolutionary effects are 
responsible for visibility of the spin pulse.

\section{Observations}
\label{sec:observations}

\begin{table*}
\centering
\caption{Log of {\it RXTE} observations. Start and end times are in
  the UT days of April 1998. Exposures are in seconds and count rates
  are in counts per second from layer 1 of detectors 0 and 1 (PCA) and from 
  cluster 0 (HEXTE). The count rate of the first observation is lower due to
  $\sim12$' source offset.} 
\vspace{12pt}
\begin{tabular}{clccrrrr}
\hline
 & & & &  \multicolumn{2}{c}{PCA} & \multicolumn {2}{c}{HEXTE} \\
No. & Obsid & Start & End & Exposure & Count rate & Exposure & Count rate \\
\hline
 1  & 30411-01-01-03S & 11.809 & 11.905 &  1968 & 256.5$\pm$0.5  & 1065  & 11.2$\pm$0.5 \\
 2  & 30411-01-02-04S & 13.078 & 13.100 &  1312  & 296.4$\pm$3.2   & 449 & 16.4$\pm$0.8 \\
 3  & 30411-01-03-00  & 16.721 & 17.194 & 14032  & 182.4$\pm$0.2  & 4483 & 10.7$\pm$0.2 \\
    & 30411-01-04-00 \\
 4  & 30411-01-05-00  & 18.131 & 18.388 & 14320  & 157.2$\pm$0.2  & 4549  & 9.1$\pm$0.2 \\
 5  & 30411-01-06-00  & 18.860 & 19.044 &  7296   & 149.4$\pm$0.2  & 2481  & 7.9$\pm$0.3 \\
 6  & 30411-01-07-00  & 20.877 & 20.964 &  4224  & 124.8$\pm$0.2  & 1451  & 6.7$\pm$0.4 \\
 7  & 30411-01-08-00  & 23.662 & 23.973 & 14512  & 105.4$\pm$0.2  & 4726  & 5.3$\pm$0.2 \\
 8  & 30411-01-09-01  & 24.672 & 24.974 & 14160  & 101.3$\pm$0.2  & 4708  & 5.3$\pm$0.2 \\
    & 30411-01-09-02 \\
 9  & 30411-01-09-03  & 25.599 & 25.904 & 14352  &  96.4$\pm$0.1  & 4724  & 5.3$\pm$0.2 \\
    & 30411-01-09-04 \\
10  & 30411-01-09-00  & 26.668 & 26.978 & 15456  &  63.9$\pm$0.1  & 5068  & 3.2$\pm$0.2 \\
11  & 30411-01-10-01  & 27.604 & 27.814 &  8896  & 33.5$\pm$0.1  & 2914  & 2.1$\pm$0.3 \\
    & 30411-01-10-02 \\
12  & 30411-01-10-00  & 29.600 & 29.790 &  8512  &  26.5$\pm$0.1 & 2303  & 2.2$\pm$0.3 \\
\hline
\end{tabular}
\label{tab:obslog}
\end{table*}


We analyse {\it RXTE} observations of 11--29 April 1998 (HEASARC
archival number P30411) outburst of the millisecond pulsar SAX
J1808.4--3658 using {\sc ftools} 5.0. We extract the PCA energy
spectra from the top layer of detectors 0 and 1 from the Standard-2
data files. A comparison with the Crab spectra shows that a 0.5 per
cent systematic error is appropriate in RXTE epoch 3 PCA observatios
for this restricted data selection between energies of 3--20 keV
(Wilson \& Done 2001). These PCA spectra are combined with the 20--150
keV data from HEXTE cluster 0. The relative normalization of the PCA
and HEXTE instruments is still uncertain, so we allow this to be an
addition free parameter in all spectral fits. Table \ref{tab:obslog}
contains the log of observations. Observation 1 has a $\sim 12$'
pointing offset, therefore the count rates are lower than in the
second observation, though the actual X-ray flux is higher.

We also extract phase-resolved energy spectra from the PCA Event mode
data files with timing resolution of 122 $\mu$s (except for the April
13 observation where we used GoodXenon mode files, with resolution of
1 $\mu$s). We have generated folded light-curves in 16 phase bins, for
each PCA channel, all layers, detectors 0 and 1. We have chosen
beginning of the phase ($\phi = 0$) at the bin with lowest 3--20 keV
count rate.  Photon arrival times were corrected for orbital movements
of the pulsar and the spacecraft. Background files were made from
Standard-2 data for exactly the same periods when our phase-resolved
spectra were extracted. Power density spectra (PDS) were extracted
from the same data files as the phase-resolved spectra, in the 3--20
keV energy range, though we have used all detectors this time. We
calculate power density spectra in the $^1/_{128}$--2048 Hz frequency
range from averaging fast Fourier transforms over 128 s data
intervals.

All spectral analysis (both phase-resolved and phase-averaged) is done
using the {\sc xspec} 11 spectral package (Arnaud 1996). The error of
each model parameter is given for a 90 per cent confidence interval.
We fix the absorption to the interstellar Galactic one in the
direction of SAX J1808.4--3658 of $1.3 \times 10^{21}$ cm$^{-2}$, and
use photoelectric cross-sections of Morrison \& McCammon (1983).

There is no good determination of the inclination of the system. Lack
of X-ray eclipses yields the upper limit of the inclination angle, $i
< 81^\circ$. On the other hand modelling of the optical companion's
multi-band photometry with a simple X-ray heated disc model suggests
$i > 63^\circ$ (Bildsten \& Chakrabarty 2001). For reflection models
we simply fix $i = 60^\circ$, a value which we will show to be
consistent with other source properties later on. We assume solar
abundances of the reflector. To convert fluxes to luminosity we use
the distance to the source of 2.5 kpc (in 't Zand et al.\ 2001).

\section{Results}
\label{sec:results}

\subsection{Outburst light-curve}
\label{sec:lightcurve}

\begin{figure}
\begin{center}
\leavevmode
\epsfxsize=7cm
\epsfbox{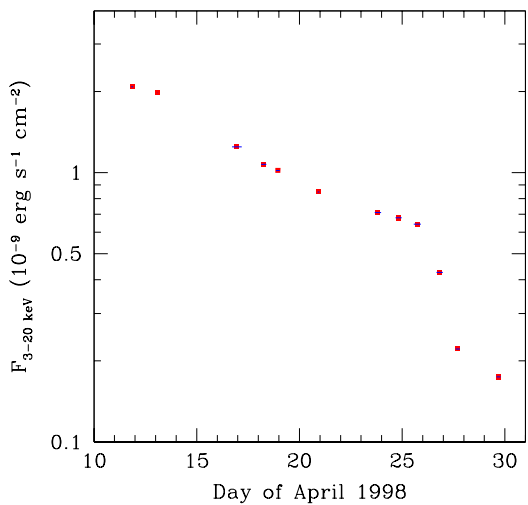}
\end{center}
\caption{Outburst of SAX J1808.4--3658: 3--20 keV unabsorbed flux of
  observations 1--12 (Table \ref{tab:obslog}).}
\label{fig:lightcurve}
\end{figure}

The 3--20 keV flux evolution after the outburst is presented in 
Fig.\ \ref{fig:lightcurve} (see also Cui, Morgan \& Titarchuk 1998; 
Gilfanov et al.\ 1998). During about 20 days after the outburst the 
flux declined by about factor ten. The bolometric unabsorbed flux 
(obtained using the model described below in Section 
\ref{sec:phase-averaged}) at the maximum is $5.1\times10^{-9}$ erg 
cm$^{-2}$ s$^{-1}$ which corresponds to bolometric luminosity of 
about $3.8\times10^{36}$ erg s$^{-1}$, which is 2.2 per cent of 
Eddington luminosity. The outburst follows a roughly exponential 
decay, with a sudden drop in the lightcurve around $26^{\rm th}$ of 
April.


\subsection{Phase-averaged energy spectra}
\label{sec:phase-averaged}

We select observation 3 with the best statistics for spectral fits.
We model this first with a power-law and exponential cutoff but the
fit is very poor, with $\chi^2$ = 351 for 82 degrees of freedom. There
are significant residuals in the PCA above $\sim$ 7 keV. We model
these as Compton reflection using angle-dependent Green's functions of
Magdziarz \& Zdziarski (1995) and a narrow Gaussian representing iron
K$\alpha$ fluorescent line.  The reflected spectrum is determined by
the solid angle subtended by the reflector, $\Omega$, inclination
angle $i$, and ionization parameter $\xi \equiv 4 \pi F_{\rm ion} / n$
(where $F_{\rm ion}$ is the 0.005-20 keV irradiating flux in a
power-law spectrum and $n$ is the density of the reflector).  We allow
for relativistic smearing of the reflected component and the line for
a given inner disc radius, $R_{\rm in}$. The fit is much better now,
$\chi^2$ = 101/78, but there is still a strong excess below $\sim$ 5
keV. Including a blackbody soft component finally gives a good fit,
with $\chi^2$ = 70.6/77.  Parameters for this best fit
phenomenological model are given in Table \ref{tab:power_law}.

\begin{table}
\centering
\caption{Best-fitting parameters of the power law plus blackbody and
  reflection model fitted to the data set 3. The reflector inclination 
  was 60$^\circ$, and we assumed solar abundances. $\xi$ and 
  $R_{\rm in}$ were frozen, while computing the parameter errors.} 
\vspace{12pt}
\begin{tabular}{lr}
\hline
Parameter & Value \\
\hline
$kT_{\rm soft}$ (keV) & $0.77\pm0.03$\\
$N_{\rm soft}$ (km$^2$) & $10.6\pm1.9$\\
$\Gamma$ & $1.82\pm{0.04}$\\
$E_f$ (keV) & $180_{-60}^{+120}$\\
$\Omega/2\pi$ & $0.05\pm{0.04}$\\
$\xi$ (erg cm s$^{-1}$) & 1900\\
$R_{\rm in} (R_g)$ & 30\\
$E_{\rm line}$ (keV) & $6.5\pm0.15$\\
$EW_{\rm line}$ (eV) & $85\pm20$\\
$\chi^2$ & 70.6/77 d.o.f. \\
\hline
\end{tabular}
\label{tab:power_law}
\end{table}

We replace the phenomenological model above with a more physically
motivated description of the spectrum.  Instead of the power law we
use an approximate solution of the Kompaneets (1956) equation
(Zdziarski, Johnson \& Magdziarz 1996) to model the thermal
Comptonization hard component in the spectrum. The model {\tt thcomp}
is parameterized by the asymptotic power-law photon index, $\Gamma$,
electron temperature, $T_e$ and blackbody seed photons temperature,
$T_{\rm seed}$. Though the seed photons temperature {\em can\/} be
independent of the soft component temperature, the simplest solution
is to equalize them. Later on we will argue that indeed the observed
blackbody is the source of seed photons. We calculate the reflected
spectrum from this continuum using a more sophisticated reflection
model where both the continuum reflection and the line are calculated
self-consistently for a given ionization state ({\.Z}ycki, Done \&
Smith 1998). The reflected spectrum (continuum and line) is
relativistically smeared for a given inner disc radius, $R_{\rm in}$.

For the soft component we use a single-temperature blackbody. Again, 
the inclusion of the soft component is {\em required} by the data: 
using just a Comptonization model and its reflection gives 
$\chi^2=174/79$. Allowing the absorption to be free reduces this 
only to $\chi^2=166/78$, whereas with a blackbody the fit is 
$\chi^2=73.6/80$. The same is if we use more conservative 1 per cent 
systematic errors in the PCA spectra instead of 0.5 per cent. A fit 
without the soft component gives $\chi^2 = 130/79$ while adding a 
blackbody reduces $\chi^2$ to 58.2/80. The lack of data below 3 keV 
in the PCA means that we see only the high-energy part of the soft 
component, so its overall spectral shape is poorly constrained. A 
multi-colour disc spectrum can equally well fit the soft excess, but 
we choose to use a blackbody as its variability implies that it 
arises from a hot spot on the neutron star surface (see section 
3.4).

Due to poor statistics, the electron temperature is poorly constrained
by the HEXTE data, particularly for later observations when the source
is fainter.  From the fits we obtain lower limits on $kT_e$ only,
typically $\ge$30 keV.  Similarly, the reflector ionisation state and
amount of relativistic smearing are also poorly constrained.  To get
the maximum possible signal--to--noise we co--add observations 3--12
and fit this total spectrum. This gives an electron temperature of
$kT_e = 90_{-30}^{+240}$ keV. The reflector ionisation state is $\xi =
1000_{-800}^{+2100}$ erg cm s$^{-1}$, and there is still only an upper
limit on the amount of relativistic smearing.  Hereafter we fix the
electron temperature and disk ionisation at these best fit values, and
fix the inner disc radius at 50$R_g$ (where $R_g \equiv GM/c^2$).

\begin{figure}
\begin{center}
\leavevmode
\epsfxsize=7cm
\epsfbox{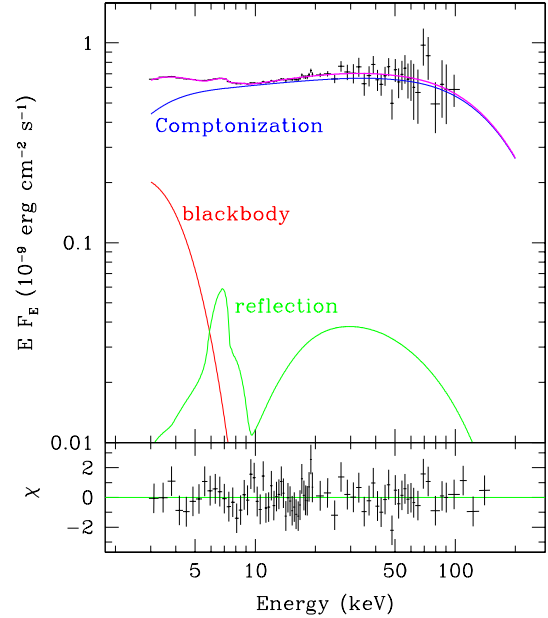}
\end{center}
\caption{Phase-averaged energy spectrum of SAX J1808.4--3658
  (observation 3) with a model consisting of a thermal Comptonization,
  blackbody and Compton reflection.}
\label{fig:spectrum}
\end{figure}

\begin{figure}
\begin{center}
\leavevmode
\epsfxsize=8.5cm \epsfbox{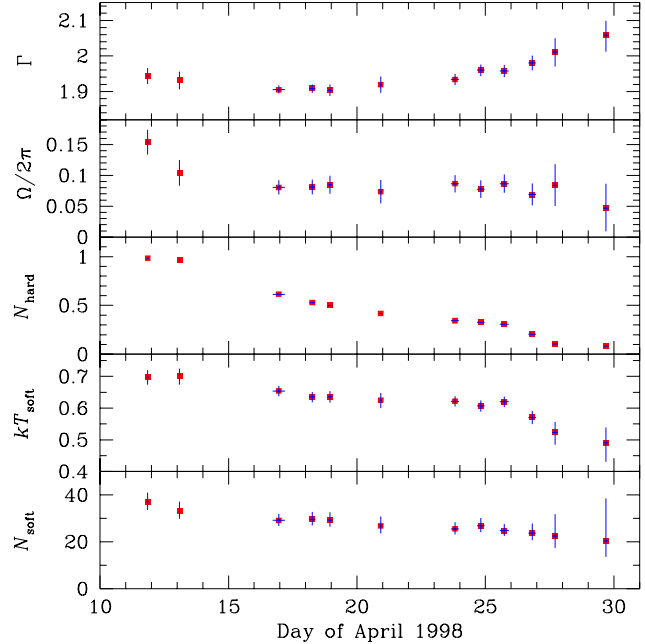}
\end{center}
\caption{Fit results of the blackbody plus thermal Comptonization and
  reflection model to the phase-averaged spectra of observations
  1--12. $\Gamma$ is the asymptotic power-law photon index, $\Omega$
  is the solid angle subtended by the reflector, $N_{\rm hard}$
  (10$^{-9}$ erg cm$^{-2}$ s$^{-1}$) is the normalization of the hard
  (Comptonization) component at 10 keV, $kT_{\rm soft}$ (keV) and
  $N_{\rm soft}$ (km$^2$) are the temperature and apparent area of
  the soft (blackbody) component, respectively.}
\label{fig:fit_params}
\end{figure}

We fit all the phase averaged spectra using this model consisting of
the thermal Comptonization, Compton reflection and the blackbody. Fit
results are presented in Fig.\ \ref{fig:fit_params}, and an example
of the spectral decomposition is shown in Fig.\ \ref{fig:spectrum}.

Despite an order of magnitude change in the observed flux
during the declining part of the outburst, the energy spectrum did not
change very much (see Gilfanov et al.\ 1998). There is a slight
softening in the hard component, starting from 24$^{\rm th}$ April
onwards. The amount of reflection is always low, $\Omega/2\pi \sim
0.08$, except for the first observation, where it was about a factor
of 2 times higher. The soft component decreased both its temperature
and apparent area, though this did not affect the overall spectral
shape very much.

\begin{figure}
\begin{center}
\leavevmode
\epsfxsize=8.5cm \epsfbox{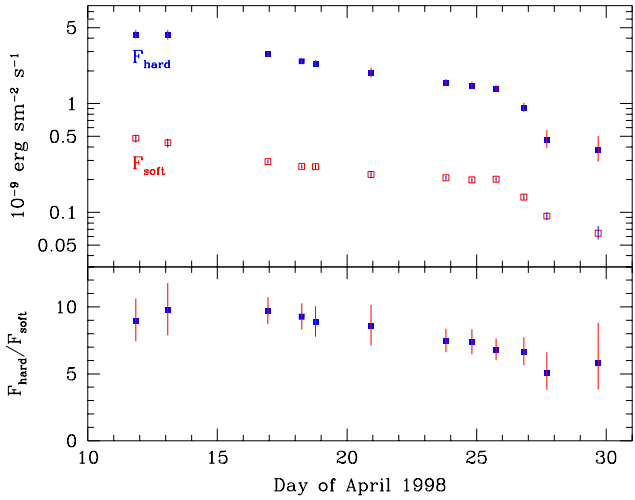}
\end{center}
\caption{Bolometric fluxes from the best-fitting model of the hard and soft 
components.}
\label{fig:fluxes}
\end{figure}

We use the unabsorbed best-fitting model (without reflection) to
compute the bolometric fluxes from the hard and soft components (Fig.\
\ref{fig:fluxes}). The ratio of $F_{\rm hard}/F_{\rm soft}$ decreased
roughly by factor two, though the errors of measurement are
substantial and, to some extent, model dependent. It is however clear
that the evolution of $F_{\rm soft}$ closely followed that of $F_{\rm
  hard}$.
Thus, both components must be closely linked in some way.

The Comptonization model we use here is only an analytical
approximation. Therefore, we check it's validity and how it affects
our fitting results, using {\tt CompPS} Comptonization code (Poutanen
\& Svensson 1996), which finds a numerical solution of the
Comptonization problem explicitly considering successive scattering
orders. We fit the co-added spectrum assuming that the Comptonizing
cloud has a shape of a cylinder with height--to--radius ratio $H/R =
1$ and the seed photons come from the bottom. We find the best-fitting
electron temperature $kT_e = 43_{-6}^{+9}$ keV and vertical optical
depth of the cylinder of $\tau = 2.7_{-0.4}^{+0.3}$. While the Compton
parameters change, the spectral shape is very similar, and the
inferred soft excess is almost identical (blackbody parameters of
$kT_{\rm soft} = 0.65\pm0.02$ keV and $N_{\rm soft} = 29\pm3$ km$^2$
with {\tt thcomp} compared to $kT_{\rm soft} = 0.68\pm0.02$ keV and
$N_{\rm soft} = 33\pm3$ km$^2$ for {\tt compPS}). Therefore, we
conclude that the detailed choice of the Comptonization model does not
affect the fit results of the soft component.

We have assumed solar abundances of the reflector, though we expect
the companion star to be highly evolved and helium rich (see e.g.\ 
Ergma \& Antipova 1999). However, the shape of the reflected spectrum
in X-rays is determined mostly by heavy elements, so variation in the
abundance of hydrogen and helium would affect the reflected spectrum
only very little (e.g.\ George \& Fabian 1991). Additionally, the
amount of reflection is rather small, so the effect of abundances on
the total spectrum is even weaker. The spectral resolution and
response uncertainties of the PCA do not allow us to investigate this 
issue in detail.

Next, we check for the presence of the Compton reflection from the
neutron star surface. Depending on the equation of state we should
expect this reflection to be red-shifted by $z$ = 0.07--0.77 (Miller,
Lamb \& Psaltis 1998). This means that the most prominent reflection
features, the iron K line and edge, can appear between 3.6 and 6.0
keV. Unfortunately, this coincides with the PCA instrumental Xenon
edge at 4.8 keV, therefore the following results should be treated
with caution. We add another reflection component in our model with
redshift allowed to be free between the above limits and fit the data.
If the neutron star surface is fairly highly ionized, with $\xi =
1000$ erg cm s$^{-1}$, we find an upper limit for $\Omega_*/2\pi <
0.012$ for a redshift range of $z$ = 0.07--0.77.  However, if the
neutron star surface is completely ionized, the reflected spectrum is
simply a featureless power law with high-energy cutoff around 30 keV
and the same spectral index as irradiating Comptonization. In such a
case we cannot disentangle the irradiating and reflected continuum, so
there are no constraints on the amplitude of the reflection from the
neutron star.

\subsection{Phase-resolved energy spectra}
\label{sec:phase-resolved}

\begin{figure}
\begin{center}
  \leavevmode \epsfxsize=8cm \epsfbox{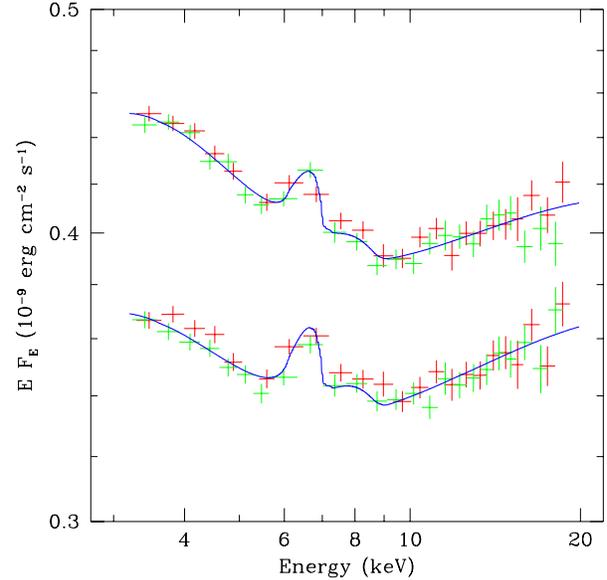}
\end{center}
\caption{Phase-resolved spectra with lowest and highest $N_{\rm hard}$, 
  corresponding to phase 0 and 0.44, respectively. The vertical axis
  is blown-up here to enhance the features of the spectra.}
\label{fig:minimax}
\end{figure}

\begin{figure}
\begin{center}
  \leavevmode \epsfxsize=8.5cm \epsfbox{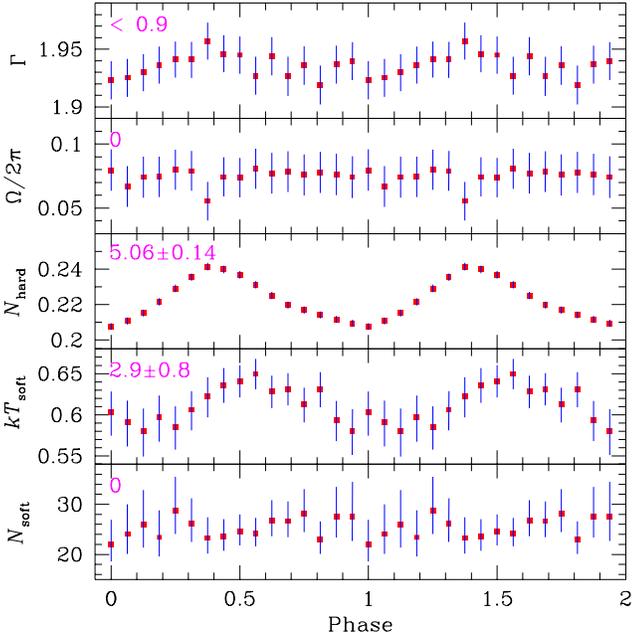}
\end{center}
\caption{Fit results of the blackbody plus thermal Comptonization and
  reflection model fitted to the phase-resolved spectra (co-added
  observations 3-12). Numbers in the panels show the intrinsic
  variance, $S$, of the given parameter (per cent). The fit parameters
  are described in the text and in the caption to the Fig.\ 
  \ref{fig:fit_params}.}
\label{fig:phase_free}
\end{figure}

The phase-resolved spectra with the lowest and highest flux are shown
in Fig.\ \ref{fig:minimax}. There is a slight softening of the
highest flux spectrum at energies below $\sim$ 5 keV, but otherwise
the spectral shape is remarkably constant.  To investigate the details
of spectral evolution with the pulse phase we fit the phase-resolved
spectra by the blackbody plus thermal Comptonization model ({\tt
  thcomp}), the same we used in Section \ref{sec:phase-averaged} to
fit the phase-averaged spectra. We fit each spectrum of the 16 phase
bins and find the fit parameters and their errors. The results are
presented in Fig.\ \ref{fig:phase_free}, with an average
$\chi^2_\nu=45.9/44$ per spectrum.

We clearly see that some parameters vary more than the others, in
particular reflection ($\Omega/2\pi$) is consistent with a constant
while $N_{\rm hard}$ exhibits strong sinusoidal pulsations. We want to
distinguish between the parameters that vary {\em significantly\/} and
those in which variability can be due to statistical noise. Therefore,
we quantify the variability of a given parameter by an {\em
  intrinsic\/} variance.  Let $x_i$ and $\sigma_i$ be a given
parameter's best-fitting value and variance, in the $i$-th phase bin
($i = 1, ..., N$; here $N = 16$).  Then, the intrinsic variance of
this parameter is estimated by
\begin{equation}
S={\sqrt{\left<\sigma^2\right>(\chi^2_\nu-1)} \over \left<x\right>},
\label{eq:s_int}
\end{equation}
where
\begin{equation}
\left<\sigma^2\right> \equiv \left({1 \over N} \sum_{i=1}^N {1 \over
  \sigma_i^2}\right)^{-1}.
\end{equation}
The reduced $\chi^2_\nu$ results from fitting a series of $x_i$ (for
$i = 1, ..., N$) by the constant. When $\chi^2_\nu < 1$ we assume $S
\equiv 0$. We estimate error of $S$ by propagating data errors,
\begin{equation}
\Delta S \equiv \sqrt{\sum_{i=1}^N \left({\partial S \over \partial
      x_i}\right)^2 \sigma_i^2} = {1 \over \left<x\right>} \sqrt{{\left< \sigma^2 \right> \over 
    N-1} {\chi^2_\nu \over \chi^2_\nu - 1}}.
\end{equation}
When $S$ is zero, then any variability is completely consistent with
having a statistical origin so that there is no significant intrinsic
variability in the parameter.  In such a case we decide that the
parameter is intrinsically constant, and fix it at its average 
value for the next series
of fits.

\begin{figure*}
\begin{center}
\leavevmode
\epsfxsize=14cm \epsfbox{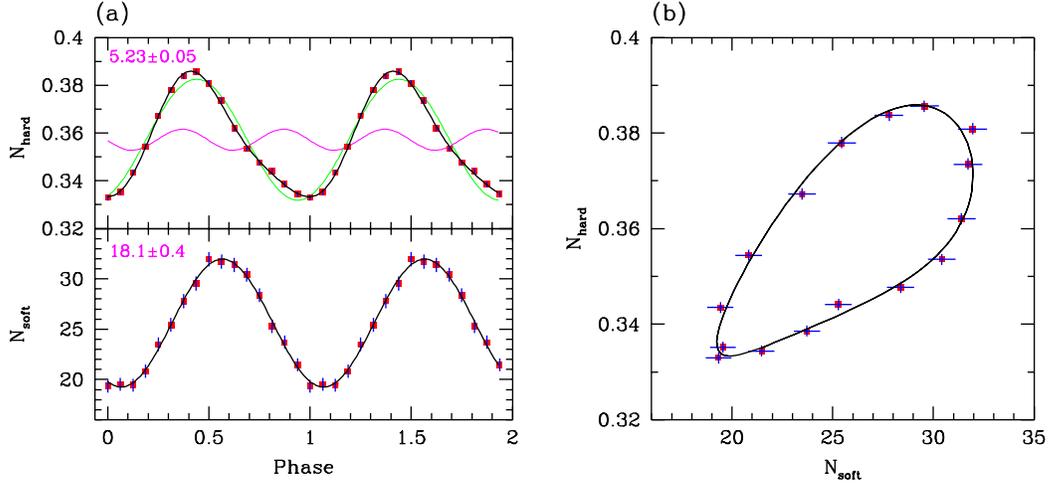}
\end{center}
\caption{Pulsation of the soft and hard components of the spectrum. 
  The co-added phase-resolved spectra were fitted by the blackbody
  plus thermal Comptonization and reflection model , where only
  $N_{\rm soft}$ and $N_{\rm hard}$ were free.  See Fig.\ 
  \ref{fig:phase_free} for the fit with other parameters free.  The
  fit parameters are described in the text and in the caption to the
  Fig.\ \ref{fig:fit_params}. The models fitted to the pulse-profiles
  are described in Section \ref{sec:phase-resolved}.}
\label{fig:phase_param}
\end{figure*}

Using this we found that only two parameters exhibited significant
intrinsic variability: $N_{\rm hard}$ and $T_{\rm soft}$ (see Fig.\
\ref{fig:phase_free}). We fixed $\Gamma$ and $\Omega/2\pi$ (but not
$N_{\rm soft}$ for a moment) at their average values and refit the
spectra (average $\chi^2_\nu=47.3/46$, so a difference of
$\Delta\chi^2=1.4$ for 2 fewer free parameters). This showed an
anti-correlation between the pulse profiles of $T_{\rm soft}$ and
$N_{\rm soft}$, which is most probably a sign of degeneracy in the
soft component parameterization. When we additionally fix $N_{\rm
  soft}$ the average $\chi^2_\nu$ rises only to 48.2/47, however
fixing $T_{\rm soft}$ instead causes significant worsening of the fits
with average $\chi^2_\nu$ = 54.1/47, which means that temperature
pulsation is preferred over normalization pulsation.  However, from
the physical point of view we should expect more variability in
apparent area (normalization) then in temperature (see discussion in
Section \ref{sec:hotspot}). We see only Wien part of the blackbody
spectrum in the PCA, so there is some degeneracy in the model, namely
in the soft component temperature, normalization and the hard
component low-energy turnover. This creates model-dependent
uncertainty. Therefore, we decide to chose the model with worse
$\chi^2$ but better physical foundation, and to fix $T_{\rm soft}$,
letting only the normalizations of the soft and the hard components
($N_{\rm hard}$ and $N_{\rm soft}$) to be free, and fit the data in
each phase bin again.  The final result is presented in Fig.\ 
\ref{fig:phase_param}$a$.

Thus, we decompose the phase resolved spectra into two independently
pulsing components: the soft and the hard one. We fit the pulse
profiles of both components by harmonic functions. The $N_{\rm hard}$
pulse profile is clearly asymmetric and cannot be fitted by a single
harmonic (we obtain $\chi^2$ = 116/13).  However, addition of another
harmonic function provides with a very good fit ($\chi^2$ = 6.0/11).
Our best-fitting function is
\begin{eqnarray} 
  \lefteqn{N_{\rm hard}(\phi) = 0.357 + 0.025 \sin[2\pi (\phi - 0.19)]}
  \nonumber \\ 
  \lefteqn{~~~+ 0.0044 \sin[2 \pi (\phi - 0.24) / 0.5)] ~ (10^{-9}
  {\rm erg s}^{-1} {\rm cm}^{-2})} 
\end{eqnarray} 
(where $\phi$ is the phase). The $N_{\rm soft}$ pulse profile is
symmetric and a single harmonic function,
\begin{equation} 
  N_{\rm soft}(\phi) = 25.6 + 6.4 \sin[2 \pi (\phi - 0.32)] ~ ({\rm km}^2), 
\end{equation} 
gives a very good fit, $\chi^2 = 3.63/13$. A second harmonic is
statistically insignificant, as it improves the fit only by
$\Delta\chi^2 = 0.05$, while taking away two degrees of freedom. The
upper limit on the second harmonic is 7 per cent of the amplitude of
the first one. We show the pulse profiles of both components and the
best-fitting functions in Fig.\ \ref{fig:phase_param}.

\begin{figure}
\begin{center}
  \leavevmode \epsfxsize=8cm \epsfbox{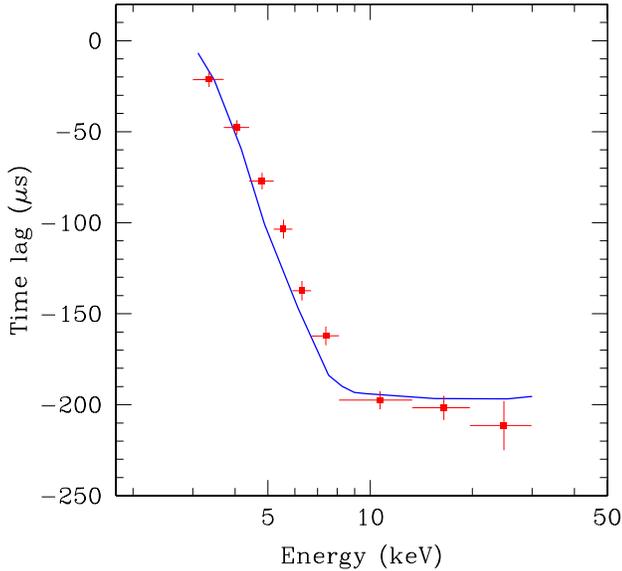}
\end{center}
\caption{Energy-dependent time lags. The data points were taken from Cui at 
  al.\ (1998) and the curve represents the prediction of the model
  with $N_{\rm hard}$ and $N_{\rm soft}$ free.}
\label{fig:lags}
\end{figure}

The pulse profiles of $N_{\rm hard}$ and $N_{\rm soft}$ are shifted 
in phase in a way that the soft component lags the hard one. This 
will manifest itself as an energy dependent time lag in which soft 
X-rays lag hard X-rays, as reported in previous work (Cui et al.\ 
1998). In Fig.\ \ref{fig:lags} we show the time lags taken from Cui 
et al.\ (1998) and the prediction of our two-component model in 
which only the normalizations $N_{\rm hard}$ and $N_{\rm soft}$ are 
free to vary. Up to about 7 keV, as the contribution from the soft 
component decreases, the soft time lag quickly increases. Above this 
energy, where the soft component is negligible, the time lag 
flattens, and shows the $\sim$ 8 per cent shift in phase (about 200 
$\mu$s) between the soft and hard components.

\begin{figure}
\begin{center}
\leavevmode
\epsfxsize=8cm \epsfbox{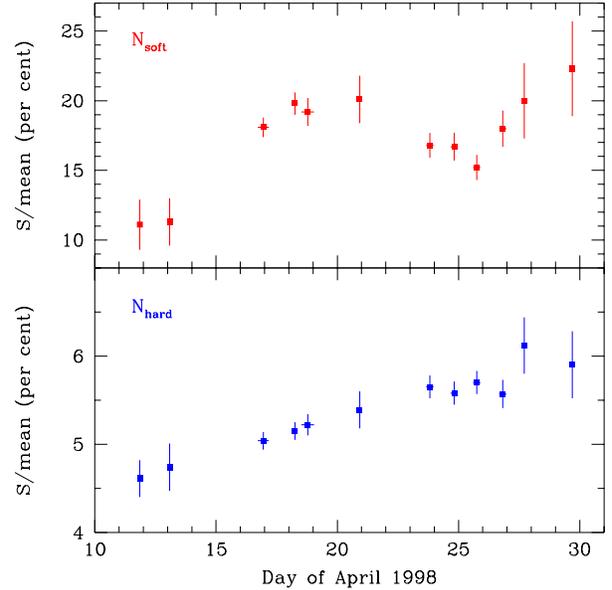}
\end{center}
\caption{Evolution of intrinsic variance (\ref{eq:s_int}) of the two
  spectral component normalizations, $N_{\rm soft}$ and $N_{\rm
    hard}$, when the data is fitted by the blackbody plus
  Comptonization and reflection model. Other model parameters were
  fixed during the fits.}
\label{fig:rms}
\end{figure}

Finally, we use the model with only $N_{\rm soft}$ and $N_{\rm hard}$
free to fit the phase-resolved spectra for each of the observation
from 1 to 12. We measure the intrinsic variance of both of the fit
parameters, and show the result in Fig.\ \ref{fig:rms}. The intrinsic
variance of $N_{\rm hard}$ increases monotonically from $\sim$ 4.5 to
$\sim$ 6 per cent during the declining part of the outburst. The
behaviour of the $N_{\rm soft}$ variance is more complex: it slowly
rises to its maximum around 18 April, then declines only to rise again
after 26 April.

Our approach contrasts with that of Ford (2000) in that we use
standard background subtraction techniques to extract the phase
resolved spectra whereas he used the phase minimum spectrum as the
persistent emission and subtracted this from the phase-resolved
spectra to get the spectrum of the pulsed component.  However, our
results clearly show that there are {\em two} pulsed components, the
hard and the soft, and that these {\em do not} vary in phase.  Ford
(2000) fit the pulsed emission by a power law, and noticed significant
phase variability of the spectral index. However, in this approach a
single power law tries to follow the overall spectral shape, which is
actually more complex.  Additionally, poor statistics of the
persistent-emission-subtracted spectra make fitting more difficult. We
have reproduced this approach but fitted these pulsed spectra with a model
consisting of a power law {\em and} a blackbody. Interestingly, the
power-law index was consistent with remaining constant with phase,
while the blackbody and power-law normalizations were shifted in
phase, as in our analysis.

\subsection{Power density spectra}
\label{sec:pds}

\begin{figure}
\begin{center}
\leavevmode
\epsfxsize=8cm \epsfbox{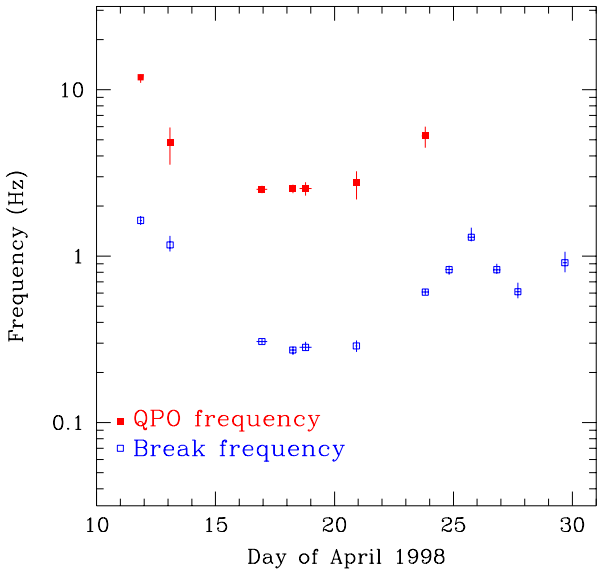}
\end{center}
\caption{Break and QPO frequencies in the power density spectra.}
\label{fig:pow_param}
\end{figure}

The properties of the power-density spectrum of SAX J1808.4--3658 were described 
in details by Wijnands \& van der Klis (1998b). Here we re-fit the PDS selected 
from the observations 1--12 (Table \ref{tab:obslog}), using similar model, in 
order to obtain break and QPO frequencies, corresponding to the exact times at 
which the data were taken. We show the results in Fig.\ \ref{fig:pow_param}, 
for completeness.

None of the two frequencies is correlated with the X-ray flux 
(Fig.\ \ref{fig:fluxes}), with any of the spectral parameters 
(Fig.\ \ref{fig:fit_params}), nor with spectral colour. While the 
spectral parameters either remain constant or change monotonically, 
both break and QPO frequencies reach their minimum around 18 April. 
After 24 April we do not detect QPO, but the break frequency 
reverses its rise around 26 April. A similar behaviour (though in 
opposite direction) can be seen in the soft component variance 
(Fig.\ \ref{fig:rms}). Thus, there is a clear anti-correlation 
between $\nu_{\rm break}$ and $S(N_{\rm soft})$.

\section{Discussion}
\label{sec:discussion}

We now have many interrelated constraints, which can be used to 
determine the geometry of the system, and how this changes as the 
outburst declines. Since both hard and soft X-rays pulsate then they 
must be affected by the spin of the neutron star. We assume that the 
underlying geometry involves a hotspot on the surface of the neutron 
star to produce the soft pulsed component, and a hard X-ray shock 
above the magnetic poles to produce the pulsed hard X-ray component. 
The constancy of the hard X-ray spectral shape with spin phase 
implies that the hard X-rays are from a single type of emission 
region, rather than being the sum of a pulsed and unpulsed component. 
The sinusoidal light-curve for the soft component as a function of 
spin phase strongly implies that we see only one hotspot. The shape 
of the hard component light-curve probably implies that we don't see 
the second shock.

More constraints come from the variability characteristics. It is commonly 
accepted that the QPO frequency is related to the inner disc radius in some way. 
The inner disc radius, together with the shock height and the neutron star 
radius, determines the solid angle subtended by the disc, and so the amount of 
reflection and seed photons from the disc. Seed photons can also arise from the 
hotspot on neutron star surface, and their contribution is determined mostly by 
the shock height.  But the spectral shape gives us observational constraints on 
the amount of reflection from the disc, and the amount of the seed photons, so 
there should be a unique solution to the geometry of the system. In practice of 
course, there are significant observational and theoretical uncertainties, but 
we use these observational constraints to develop a consistent geometrical 
picture for the emission from SAX J1808.4--3658.

\subsection{Emission mechanisms}

The 3--200 keV spectrum {\em requires} a two component model. The soft
component is consistent with a blackbody at $\sim$ 0.5--0.7 keV. The
hard component can be described by thermal Comptonization of these
blackbody seed photons in plasma of optical depth of $\sim$ 2 and
temperature of $\sim$ 40 keV. The soft and hard component fluxes are
very closely correlated (see Fig.\ \ref{fig:fluxes}). Thus the soft
component is most probably from reprocessing of the hard component. We
further discuss relation between both components in Appendix
\ref{sec:energy_balance}.

\subsection{Hotspot}
\label{sec:hotspot}

We consider the source of the soft component as a simple circular
hotspot on the surface of the neutron star. We take into account
Doppler shifts and geometry in the rotating system of the neutron
star. We demonstrate, how in the frame of this simple model the data
can constrain inclination angle of the system and the size of the
hotspot.

The pulse profile of the soft component is consistent with a single
sinusoid.  This implies that we {\em do not\/} see the second spot,
since it would have given rise to a secondary maximum in the profile.
This, in turn, yields a constrain on the spot radius, \begin{equation}
  {R_{\rm spot} \over R_*} < {\pi \over 2} - i - \delta,
  \label{eq:rhs} \end{equation} where $i$ is the inclination angle of
the observer from the rotation axis, $\delta$ is the angle from the
rotation axis to the spot at the magnetic pole and $R_*$ is the
neutron star radius.

The inferred luminosity of a spot depends on both its temperature and
projected area. Both of these change as a function of phase. The Doppler shift 
gives a ratio between
the maximum and minimum temperature of
\begin{equation}
{T_{\rm max} \over T_{\rm min}} \sim {1 + \beta_{\rm eq}
\sin \delta \sin i \over 1 - \beta_{\rm eq} \sin \delta \sin i},
\label{eq:t_maxmin}
\end{equation}
where $\beta_{\rm eq} \equiv v_{\rm eq}/c \sim 0.08$ is the equatorial
rotation speed of the neutron star. The ratio of $T_{\rm max}/T_{\rm
  min}$ is only weakly affected by the spot size. The
maximum--to--minimum ratio of the projected area is
\begin{equation}
{A_{\rm max} \over A_{\rm min}} = {\cos (i-\delta) \over \cos(i+\delta)},
\label{eq:area}
\end{equation}
where the maximum area is offset in phase by $90^\circ$ from the
maximum Doppler shift. This formula is exact even for a
large circular spot, again assuming that we do not see the spot from
the second pole.

When the temperature of the soft component is a free parameter in the
fits to the phase-resolved spectra (Section \ref{sec:phase-resolved})
then it yields large modulation in this temperature, with $T_{\rm
  max}/T_{\rm min} = 1.12$ (see Fig.\ \ref{fig:phase_free}). However,
physically we expect a rather small variation in temperature from the
Doppler shift, since the offset angle between the magnetic spin and
the rotation axis is probably small (Ruderman 1991).  Assuming an
inclination angle of $i=60^\circ$ and offset angle of
$\delta=10^\circ$ we find $T_{\rm max}/T_{\rm min} = 1.02$, much less
then observed. Furthermore, this would give an inferred luminosity
modulation of about 1.08, much less than the observed
minimum--to--maximum ratio of 1.7.  Therefore, Doppler effects alone
are {\em unlikely\/} to be the predominant cause of the variability,
contrary to Ford (2000).

Conversely, for these parameters the projected area variability
results in a maximum--to--minimum luminosity ratio of 1.9. Thus it
seems likely that the majority of the luminosity change should come
from the change in projected area while the temperature remains mostly
constant.

\begin{figure}
\begin{center}
\leavevmode
\epsfxsize=8cm \epsfbox{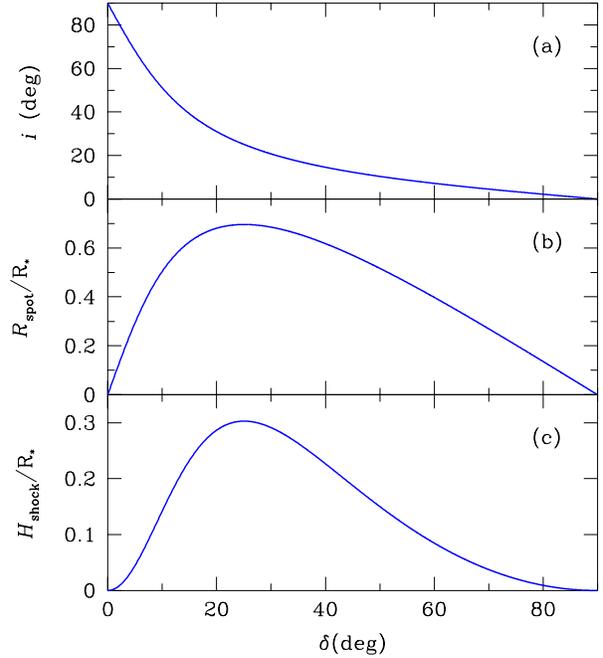}
\end{center}
\caption{$(a)$ Solution of the Equation (\ref{eq:area}),
  assuming maximum--to--minimum area amplitude of the soft component
  $A_{\rm max}/A_{\rm min} = 1.52$. $(b)$ The upper limit on the
  radius of the hotspot, from Equations (\ref{eq:rhs}) and
  (\ref{eq:area}).  $(c)$ The upper limit on the shock height from
  Equations (\ref{eq:hsh}) and (\ref{eq:area}).}
\label{fig:iarh}
\end{figure}

To check consistency of these predictions with the data, we fit the
phase-resolved spectra leaving only $N_{\rm hard}$ and $N_{\rm soft}$
as free parameters, but additionally enforcing sinusoidal variation of
$T_{\rm soft}$ with maximum--to--minimum ratio of 1.02, as predicted
using Equation (\ref{eq:t_maxmin}). As a result, maximum--to--minimum
ratio of $N_{\rm soft}$ is 1.56. We use these numbers to constrain the
inclination and offset angles. We solve Equation (\ref{eq:area}) with
$A_{\rm max}/A_{\rm min} = 1.56$.  This gives us one--to--one relation
between $i$ and $\delta$ (Fig.\ \ref{fig:iarh}$a$). For the
inclination angle of $i = 60^\circ$ the corresponding offset is
$\delta = 7^\circ$. Equations (\ref{eq:rhs}) and (\ref{eq:area}) then
give an upper limit for the hotspot radius (Fig.\ \ref{fig:iarh}$b$).
We find $R_{\rm spot} < 0.4R_*$, or $R_{\rm spot} < 4$ km, assuming
$R_*$ of 10 km.

The hotspot emission gives us an independent way to estimate its size.
From the blackbody normalization we estimate its average projected
area as $\sim$ 30 km$^2$ (see Fig.\ \ref{fig:fit_params}). This is a
lower limit as some of the hotspot could be obscured by the shock, and
we have not included gravitational redshift. Nonetheless, for an
inclination of $60^\circ$ this yields $R_{\rm spot} \ga 4$ km. Given
that we then have both an upper and lower limit for the spot size of
$\sim 4$ km, we take this as our estimate of the spot size.

For the maximum possible inclination angle of $i = 80^\circ$ the required offset 
is small, $\delta = 2^\circ$. This yields in turn $R_{\rm spot} < 0.1R_*$, which 
is not consistent with our lower limit on the hotspot size from the 
normalization of the blackbody. Therefore, we find the inclination angle of $i = 
60^\circ$ the most likely.

Our discussion neglects general relativistic effects. These will 
change the quantitative results, but for our model then the effects 
of light bending are not particularly strong. Full relativistic 
calculations by Psaltis, {\"O}zel \& DeDeo (2000) show that the 
inferred area of the spot is within 10 per cent of its intrinsic area 
for our model geometry, where the spot is large and not eclipsed. 
Thus the qualitative constraints derived here on the geometry should 
be valid.

\subsection{Shock}
\label{sec:shock}

We consider the source of the hard component as an X-ray emitting shock in 
the accretion column. We again take into account Doppler shifts and geometry.

The hard component pulse profile, like the soft one, lacks the
secondary maximum, which implies that we {\em do not\/} see the second
shock. This immediately yields strong constraints on the shock height:
\begin{equation}
{H_{\rm shock} \over R_*} < {1 \over \sin(i + \delta)} - 1.
\label{eq:hsh}
\end{equation}
The solution to this equation, solved simultaneously with the
offset-inclination relation (\ref{eq:area}) derived above is shown in
Fig.\ \ref{fig:iarh}$c$. For a small offset angle of $\delta =
7^\circ$ the shock must be short, $H_{\rm shock} < 0.1R_*$.

The hard X-ray spectrum can be described as a power law with photon
spectral index $\Gamma \approx 2$. We can expect it to be almost
isotropic. Then, in the non-relativistic limit the Doppler shift gives
the ratio of maximum--to--minimum observed hard flux, as
\begin{equation}
{F_{\rm max} \over F_{\rm min}} = \left({1 - \beta_{\rm eq}
\sin i \sin \delta \over 1 + \beta_{\rm eq} \sin i \sin \delta}\right)^{-2-\Gamma}.
\label{eq:hard_amp}
\end{equation}
For $\delta = 7^\circ$ and $i = 60^\circ$ this ratio is 1.07, while the observed 
maximum--to--minimum is 1.16. An additional source of variability could be 
reflection from the neutron star surface and/or anisotropy of the shock 
emission. The reflection from the fully ionized surface would be spectrally 
indistinguishable from the irradiating continuum and so is consistent with our 
spectral fits (see Appendix \ref{sec:ns_reflection}). For the above inclination 
and offset angles the reflection component should have maximum--to--minimum flux 
ratio of about 1.2.


{\em Unlike\/} the soft component pulse profile, the hard one is not 
symmetric: it is skewed. In Section \ref{sec:phase-resolved} we have 
fitted this profile with two sinusoids: a dominating one with the 
spin frequency and a weaker one, with double-spin frequency. This 
can be explained if the cross-section of the shock is elongated, 
with bi-axial symmetry, which would create a double-spin frequency 
on top of the main, single-spin frequency pulse profile. In reality, 
magnetic-field induced column accretion onto the neutron star 
surface might create a latitudaly-elongated, crescent-shaped shock 
cross-section.

\subsection{Phase shift}
\label{sec:phase_shift}

There are two main sources of phase-dependent variability: Doppler
shifts and variation in the projected area.  As the spot goes around
the spin pole, the projected area is maximized when the spot is
closest to the observer, while the Doppler shifts are maximized when
the spot is midway between its closest and furthest points, as this is
where its velocity along the line of sight is largest.

Both Doppler and projected area effects should affect the source
spectrum. However, the spin pulse variability of the soft component is
probably dominated by the projected area terms (see Section
\ref{sec:hotspot}). If the hard component were dominated by Doppler
shift then the two components should be shifted in phase by
90$^\circ$. However, should be some projected area effects in the hard
emission, both from observing the translucent column at differing
angles and from the presence of highly ionized reflected continuum
emission (see Section \ref{sec:shock}). This means that the hard
spectrum varies as the weighted sum of these components, one which
peaks as the Doppler emission and one which peaks as the effective
area. Thus the maximum of the hard emission moves closer to the soft
emission. While detailed modelling depends on the unknown geometry of
the accretion column, its seems possible that this can result in the
observed phase lag of $\sim 50^\circ$ (Fig.\ \ref{fig:phase_param}).

This is a very different interpretation to the observed energy
dependent phase lag than that of Ford (2000). Ford (2000) used only a
single blackbody component at $kT\sim 0.6$ keV as the pulsing spectrum
This does give energy dependent phase lags, but of course contributes
very little to the hard spectrum. Although he does get 10 per cent
variability at 20 keV, this is measured with respect to the blackbody
rather than to the total spectrum. Since the blackbody flux
contributes only a tiny fraction ($5\times 10^{-9}$) of the total flux
at 20 keV, his 10 per cent variation of the blackbody flux predicts
only a $5\times 10^{-8}$ per cent modulation at 20 keV, so cannot
explain the observed phase lags in the total spectrum.  By contrast,
our model can explain the observed time lags, as shown in Fig.\
\ref{fig:lags}. In particular, it predicts flattening of the time lag
above $\sim$ 7 keV, where the soft component is negligible. The key
assumption required by our model is its two-component nature. There is
the soft and hard spectral component and they pulsate {\em shifted in
  phase}. Whatever is the actual geometry, this phase shift is
required to see the time lags.

\begin{figure*}
\begin{center}
\leavevmode
\epsfxsize=17cm \epsfbox{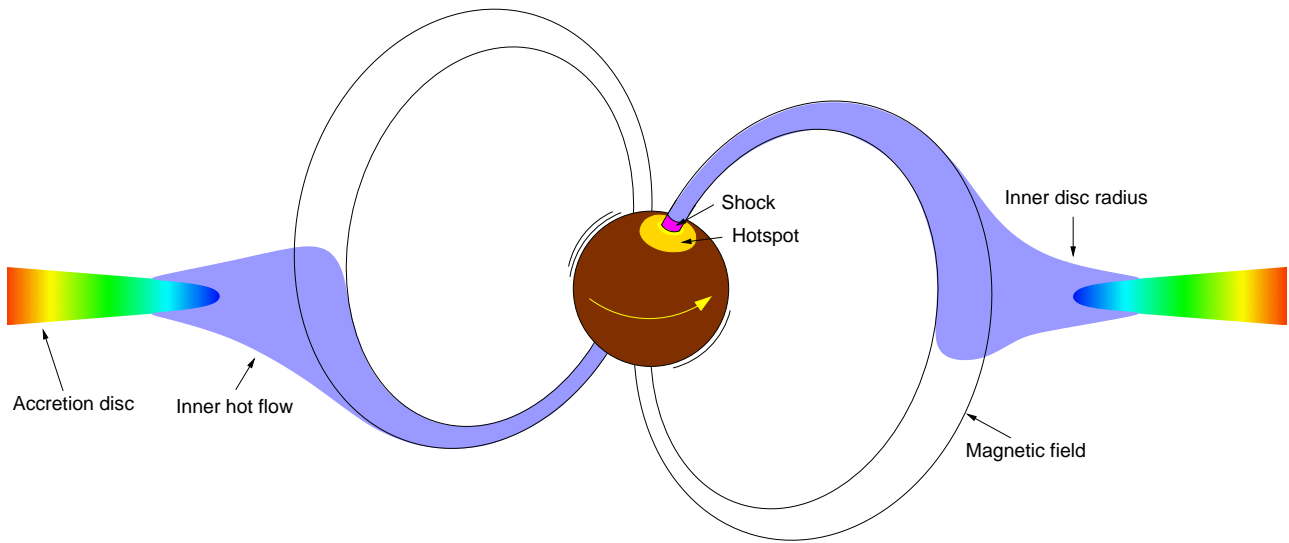}
\end{center}
\caption{Sketch of the millisecond pulsar geometry. The accretion
  disc truncates at $\sim$20--40 $R_g$ into an optically thin flow,
  which connects onto the magnetic field. The in-falling material
  creates a short shock -- the source of hard X-ray emission.
  Further re-processing of the flow creates the hotspot on the
  surface, which feeds Comptonization in the shock with the seed
  photons.}
\label{fig:ns}
\end{figure*}

\subsection{Long Term variability}

The 2--50 keV luminosity of SAX J1808.3--3658 changed by an order of 
magnitude during the decline from outburst, yet the spectral shape 
remained remarkably constant. However, neutron star binaries in 
general show a dramatic transition from hard--to--soft at a 2--50 keV 
luminosity $\sim$ 0.01 of $L_{\rm Edd}$ (Ford et al.\ 2000), 
corresponding to a bolometric luminosity of a few per cent of 
Eddington. This spectral switch is startlingly like that seen in GBHC 
at low bolometric luminosities (e.g. Nowak 1995). This similarity 
makes it likely that the underlying mechanism for the transition is 
the same irrespective of the nature of the compact object, so 
implying a connection to the accretion flow itself rather than to 
anything to do with a magnetosphere or solid surface.

Assuming the distance of 2.5 kpc (in 't Zand et al.\ 2001) the 2--50
keV flux during the outburst changed from 0.01 to 0.001 of $L_{\rm
  Edd}$ (where $L_{\rm Edd} = 2.5\times10^{38}$ erg s$^{-1}$), so it
is likely that SAX J1808.3--3658 never became quite bright enough to
change from the hard (island) state to the soft (banana) state.  The
reason for the low peak outburst luminosity is easy to find. The
binary is a very small system with an orbital period of only 2.1 hours
(Chakrabarty \& Morgan 1998), so the disc outer radius (set by tidal
truncation) is also small ($R_{\rm out} \sim 2\times10^{10}$ cm). This
limits the total disc mass $M_{\rm disc}\propto R_{\rm out}^3$ which
can build up in quiescence, so when the outburst is triggered and all
this mass flows in on the viscous time-scale $t_{\em visc}\propto
R_{\rm out}$, then the mass accretion rate onto the central object is
$M_{\rm disc}/t_{\rm visc}\propto R_{\rm out}^2$ (King \& Ritter 1998;
Shahbaz, Charles \& King 1998).

As the outburst declines there is a rather steep drop in the
light-curve, after 26$^{\rm th}$ of April (Fig.\
\ref{fig:lightcurve}). This has been identified with the
centrifugal barrier (Gilfanov et al.\ 1998). However, pulsed 
emission is {\em always\/} seen, so the magnetic field is not 
inhibiting accretion. Given that the system is transient we
suggest instead that the drop is associated with the onset of
the cooling wave, switching the disc back into quiescence
(e.g. King \& Ritter 1998; Dubus, Hameury \& Lasota 2001).

\subsection{Evolution of variance}

The variance of the soft and hard components change in a rather
different way as a function of time (see Fig.\ \ref{fig:rms}). The
hard variance monotonically increases, while the soft variance follows
the inner disc radius (assuming that this relates to the QPO
frequency).  If the disc is truncated by the magnetic field then the
point at which the material connects on is the same as the inner disc
radius. A larger disc radius means that the material connects onto
field lines which are closer to the magnetic pole, so the projected
area effects decrease, so the soft variance decreases (e.g. Frank,
King \& Raine 1992).  This is exactly {\em opposite} to the observed
behaviour (Figs.\ \ref{fig:rms} and \ref{fig:pow_param}), strongly
suggesting that the inner disc radius is truncated by some means other
than that of the magnetic field. Since other atolls (and galactic
black holes) show similar QPO behaviour then the mechanism
must be an intrinsic property of the accretion flow.

If the optically thick disc truncates at some radius into an inner, optically 
thin flow (e.g.\ R{\'o}{\.z}a{\'n}ska 1999; Meyer, Liu \& Meyer-Hofmeister 
2000), then the point at which this optically thin flow connects onto the 
magnetic field is determined by the ratio of its ram pressure to the magnetic 
pressure (e.g.\ Frank, King \& Raine 1992). But if the inner disc retreat is 
connected with increased evaporation into the hot flow (R{\'o}{\.z}a{\'n}ska 
1999) then its density, and so its ram pressure, will be inversely related to 
the inner disc radius. As the inner edge of the optically disc recedes then the 
ram pressure of the hot flow increases so it penetrates further into the 
magnetosphere before connecting onto the field lines. The material then 
collimates onto a spot which is further from the magnetic pole, increasing the 
effective angle between the spot and the spin axis, so increasing the soft 
variability.

The hard component should then also increase its variance. However, as
the mass accretion rate goes down, the shock optical depth probably
decreases, while its height probably increases. The amount of variance
apparently depends on the geometry and optical depth of the shock. In
principle it is possible that these effects could contribute to the
observed variance more then the position of the accretion column, and
so explain the observed monotonic increase in variance.

\subsection{Geometry}

A geometry which is consistent with all the above constraints is one
in which a neutron star with magnetic field $\sim 10^8$ G accretes
from a disc which evaporates into an inner hot flow at some radius
which is set by the overall evolution of the disc.  This inner hot
flow falls inward until its ram pressure is comparable with that of
the magnetic field, after which point it is collimated by the field.
It then free falls onto the magnetic poles, forming an accretion
column (see Fig.\ \ref{fig:ns}).

All the other source properties are consistent with this geometry (see
Appendices). The seed photons for Comptonization in the shock are
probably from the surface of the neutron star, while the disc emission
is negligible (Appendix \ref{sec:seed_photons}). Energy balance shows
that these seed photons cannot be only due to irradiation of the
surface by the hard X-rays from the shock. Either reflection from the
surface and/or re-processing of the kinetic energy of the in-falling
material is necessary to support the sufficient amount of the seed
photons (Appendix \ref{sec:energy_balance}). The reflection of the
hard X-rays from the neutron star surface, though not detected, can be
present in the data as a featureless cutoff power law (Appendix
\ref{sec:ns_reflection}). And finally, we applied one of the QPO
models to the data and estimated the inner disc radius to vary between
$\sim$ 15 and $\sim 45 R_g$, consistent with the observed amount of
the disc reflection (Appendix \ref{sec:disc}).

\subsection{The difference between SAX J1808.4--3658 and other atoll systems}

We speculate that the true distinction between this and other LMXBs
is purely an evolutionary effect. SAX J1808.4--3658 is highly evolved, the
companion star is mostly evaporated/accreted. The mass transfer rate
is very low, and has been low for a long time (King 2000). Models of
accretion onto a magnetized neutron star suggest that the magnetic
fields of up to $\sim 10^{11}$ G) are buried by inflow where the
accretion rate is a few per cent of Eddington. The field can diffuse
back out, but only on time-scales of 100--1000 years (Cumming, Zweibel
\& Bildsten 2001). Only systems where the time averaged mass transfer
rate has been lower than this for a few thousand years can have a
magnetic field which is high enough to collimate the accretion flow.

\section{Conclusions}

We analyse spectra from SAX J1808.4--3658 from the whole {\it RXTE\/}
campaign of 1998. This follows the April 1998 outburst as the
luminosity declines by a factor of over 10. We fit physically
motivated models to the phase-averaged spectrum from each observation,
showing that the continuum spectrum can be described by thermal
Compton scattering of soft seed photons. This spectrum is reflected
from the accretion disc, producing a weak reflected continuum and iron
K$\alpha$ line as has been seen in other hard state (low mass
accretion rate) LMXBs (Yoshida et al.\ 1993; Barret et al.\ 2000).
However, there is also an additional component in the spectrum which
can be fit by a blackbody. In phase-resolved spectroscopy it is this
component which is most variable. The intensity of the soft component
changes as the outburst progresses in the way expected for it to be
produced by reprocessing of in-falling material energy on the neutron
star surface. We show that the spectrum is consistent with a geometry
where the disc is truncated at fairly large radii, the hard X-ray
emission is from a shocked accretion column where the magnetically
collimated flow impacts onto the neutron star surface, and the
blackbody component is from the hotspot on the neutron star surface
(see Fig.\ \ref{fig:ns})

The disc cannot extend down to the last stable orbit as plainly the
inner flow is collimated by the magnetic field. The truncation radius
of the optically thick disc can be estimated by the QPO and break
frequency seen in the variability power spectra assuming that these
correspond to vertical perturbations at the inner disc edge (Psaltis
\& Norman 2001). The amount of reflection can also give an independent
estimate for the inner disc edge. The two results are consistent,
although the uncertainties are large, supporting the relativistic
orbits interpretation of the QPOs. The inner radius evolution (as
measured by the QPO frequency) is not at all consistent with the
magnetic field truncating the disc. It seems much more likely that the
accretion disc truncates spontaneously, and that the resulting inner
hot accretion flow is then collimated by the magnetic field at rather
smaller radii.

This has obvious application to other relativistic disc accreting
systems, both neutron stars and black holes. These have very similar
variability power spectra at low mass accretion rates to that of the
millisecond pulsar. If the QPO/break frequency in the millisecond
pulsar indicates the inner disc radius then it also shows that the
disc is generally truncated in all the other disc accreting systems at
low mass accretion rate. Given that we argue above that the disc in
SAX J1808.4--3658 spontaneously truncates, then it seems likely that
the same mechanism (whatever that is) operates generically in all low
mass accretion rate discs.

Finally, we speculate that the reason SAX J1808.4--3658 is currently
unique in showing spin pulsations is due to its evolutionary state.
All the well known LMXB neutron star systems have a higher time
averaged mass accretion rate which can bury the magnetic field below
the surface.

\section{Summary}

Summarizing the results from this paper, we find as follows:
\begin{itemize}
\item The X-ray spectrum of SAX J1808.4--3658 consists of the two
  components.
\item The soft component is consistent with the blackbody. We find
  that most likely this emission originates from the hotspot on the
  neutron star surface, with radius $\sim 4$~km.
\item The hard component is consistent with Comptonization of the seed
  photons from the hotspot in a plasma of intermediate optical depth
  and temperature of $\sim 40$~keV. This is probably a short ($\la 0.1
  R_*$) shock in the accretion column where the accreting material
  collimated by the magnetic field impacts onto the neutron star
  surface.
\item The hard component includes also weak Compton reflection from
  the accretion disc. The amount of reflection implies that the disc
  is truncated at fairly large radii (20--40 $R_g$), consistent with
  the lack of relativistic smearing of the spectral features.
\item Both components pulsate shifted in phase, while the neutron star
  rotates. This creates the energy-dependent time lags in which the
  soft photons lag the hard ones.
\item The inner disc radius derived from the break and QPO frequencies
  is consistent with the inner disc radius derived from the solid
  angle subtended by the reflected spectrum. It evolves independently
  of the instantaneous accretion rate, implying that the inner disc
  radius is independent of and larger then the magnetospheric radius.
  The magnetic field does not truncate the disc in SAX J1808.4--3658.
  Instead, the inner disc radius is set by some much longer time-scale
  process, most probably connected to the overall evolution of the
  accretion disc.
\item The disc truncation mechanism is probably generic to low mass
  accretion rate flows both in disc accreting neutron stars and black
  hole systems.
\item We do not detect Compton reflection from the neutron star
  surface, though we predict that it should be present in the X-ray
  spectrum. This would give an unambiguous observational measure of
  $M/R$ if there is any iron on the neutron star surface which is not
  completely ionized.
\end{itemize}

\section*{Acknowledgements}

This research has been supported in part by the Polish KBN grants
2P03D00514, 2P03D00614, by the Foundation for Polish Science
fellowship 15,4/99 and by a Polish-French exchange program.


\appendix

\section{Seed photons}
\label{sec:seed_photons}

Here we quickly discuss the possible sources of the seed photons for 
Comptonization. Let us consider a Comptonizing region eleveted at 
height $h$ above the pole of the neutron star of radius $R_*$. It 
can receive the seed photons from the accretion disc and from the 
hotspot on the surface. We consider the Shakura-Sunyaev disc 
extending from $R_{\rm in}$ to infinity, around the neutron star of 
mass $M$ and with accretion rate $\dot{M}$. The circular hotspot of 
radius $R_{\rm spot}$ and temperature $T_{\rm spot}$ is located on the 
spin pole (for these estimations only). We assume $M$ = 1.4M$_\odot$ 
and $\dot{M} = 5\times10^{16}$ g s$^{-1}$.

The energy density of the photons from the disc in the Comptonizing 
region is \begin{equation} u_{\rm disc} = {3 G M \dot{M} \over 4 c 
\pi (R_* + h)^3} \left[ {X \over R_{\rm in}} + {R_{\rm in} \over X} 
- 2\right], \end{equation} where $X = [(R_* + h)^2 + R_{\rm 
in}^2]^{1/2}$. The energy density of the photons from the hotspot in 
the Comptonizing region is \begin{equation} u_{\rm spot} = {2 \sigma 
T_{\rm spot}^4 \over c} \left[1 - {h \over \left(h^2 + R_{\rm 
spot}^2\right)^{1/2}} \right]. \end{equation} For a short shock, $h 
= 0.1R_*$, the hotspot of radius $0.4R_*$ and the inner disc radius 
of $30R_g$, $u_{\rm spot}/u_{\rm disc} \approx 900$. The two 
quantities are comparable only if the shock is tall ($h \ga 2R_*$), 
which we find unlikely (see Section \ref{sec:shock}).  

The energy density of the self-absorbed synchrotron seed photons in 
plasma of temperature $\theta \equiv kT_e /m_e c^2$ and magnetic 
field $B = B_8 10^8$G is \begin{equation} u_{\rm magn} = {8 \pi 
\nu_t^3\over 3 c^3} \, kT_e, \end{equation} where $\nu_t \approx 
9.8\times10^{16} \theta^{0.95} \tau^{0.05} B_8^{0.91}$ Hz is the 
turnover frequency (Wardzi{\'n}ski \& Zdziarski 2000). For $\theta$ 
= 0.1, $\tau = 1$ and $B = 10^8$ G we find $u_{\rm spot}/u_{\rm 
magn} \approx 400$. Therefore, we conclude that the dominating 
source of the seed photons for Comptonization is the hotspot on the 
neutron star surface, which is at the same time a source of the soft 
component (blackbody) seen in the X-ray spectrum.

\section{Energy balance}
\label{sec:energy_balance}

Let us consider a source of the hard component located above the neutron star 
surface and irradiating it. A fraction of the irradiating luminosity is then 
re-processed and re-emitted as seed photons for Comptonization. Here we check 
whether our assumed geometry is consistent with the energy balance in such a 
Comptonizing closed cycle.

The luminosity of the hard component is $L_{\rm hard}$. A fraction $d
\times \Omega_*/2\pi$ of it is intercepted by the surface, where
$\Omega_*$ is the solid angle subtended by the surface to the hard
X-ray source, and $d$ accounts for intrinsic anisotropy of
Comptonization.  The surface has albedo $A$, so the power in the
re-processed photons is $L_{\rm irr} = d (1 - A) (\Omega_*/2\pi)
L_{\rm hard}$. On the other hand the Comptonizing cloud amplifies the
seed photons, giving rise to the hard component luminosity, $L_{\rm
  hard} = C (\Omega_X/2\pi) L_{\rm seed}$, where $\Omega_X$ is the
solid angle subtended by the hard X-ray source to the hotspot and $C$
is Compton amplification factor. Hence,
\begin{equation}
{L_{\rm irr} \over L_{\rm seed}} = d C (1- A) {\Omega_X \over 2\pi} 
{\Omega_* \over 2\pi}.
\end{equation}
The maximum albedo here is $A \sim 0.8$ because the hard $\ge 30$ keV
photons are Compton down-scattered and cannot be completely reflected.
However, there is also a minimum albedo of $A \sim 0.7$ since the
observed temperature is $\sim$ 0.5--0.7 keV, and so the material is
probably highly ionized from collisional ionization.

We have performed Monte Carlo simulations of the observed spectrum and
found $d = 0.57$ and $C = 3.6$ (in the slab geometry).  Then, the
photons re-processed from irradiation can account only for about $0.4
(\Omega_X/2\pi) (\Omega_*/2\pi)$ of the seed photons. Both solid
angles are less than $2\pi$ (though not much less for a short shock),
so 0.4 constitutes a firm upper limit on $L_{\rm irr}/L_{\rm seed}$.
Therefore, there must be an additional source of the seed photons.
Since they must come from the neutron star surface (Appendix
\ref{sec:seed_photons}), we sugguest the following two possibilities.

First, the additional seed photons can originate from reflection from
the neutron star surface (see Appendix \ref{sec:ns_reflection}). Due
to large albedo the luminosity carried by the photons Compton
reflected from the surface is roughly four times luminosity in the
re-processed photons. Second, the in-falling material kinetic energy
can be directly re-processed into heating of the hotspot, additionally
to the irradiation, and this process can give rise to the blackbody
seed photons.

We note that an anisotropy of the hard X-ray source (e.g.\ due to bulk motion 
Comptonization) cannot boost the re-processed emission enough to reach the 
equilibrium. Even if all of the hard X-ray luminosity was directed towards the 
neutron star ($d = 1$), then $L_{\rm irr}/L_{\rm seed}$ would be still 
significantly less then 1.

\section{Reflection from the neutron star surface}
\label{sec:ns_reflection}

In Section \ref{sec:phase-averaged} we have derived an upper limit on
the reflection from a mildly ionized star, $\Omega_*/2\pi < 0.012$.
Such a small solid angle subtended by the star would require an
irradiating source at a distance $\sim 10R_*$ from the surface,
implying a {\em very tall\/} shock, which we find unlikely. A more
likely short shock of $H_{\rm shock} = 0.1R_*$ subtends an angle
$\Omega_*/2\pi = 0.6$. This can be in agreement with the observations
only if the reflected spectrum is featureless, which means that either
the reflecting spectrum is completely ionized or the heavy elements,
including iron, sank under the surface in the strong gravitational
field, so the surface could act as a mirror. The component reflected
from the surface would then contribute to the total spectrum as a
power-law of the same index as the irradiating Comptonization and a
high-energy cutoff above $\sim$ 30 keV. Such a component would be
impossible to disentangle from the Comptonization in {\it RXTE\/}
data.

\section{Accretion disc geometry}
\label{sec:disc}

Here we consider a particular QPO model and check whever the results obtained 
from it are consistent with the our assumed geometry of SAX J1808.4--3658. If 
the (low frequency) QPO is connected to the precession time-scale of a vertical 
perturbation in the disc (Stella \& Vietri 1998; Psaltis \& Norman 2001) then 
its frequency is uniquely linked to some transition radius (we assume it is the 
inner radius $r_{\rm in} \equiv R_{\rm in}/R_g$) in the disc, as (Psaltis \& 
Norman 2001)
\begin{equation}
\nu_{\rm QPO} \approx 2\times10^5 {1 \over mr_{\rm in}^{3/2}} {1 -
  \sqrt {1-{4a_* r_{\rm in}^{-3/2}}} \over \sqrt{1 + {a_* r_{\rm
        in}^{-3/2}}}} \ {\rm Hz},
\label{eq:qpo}
\end{equation}
where $m \equiv M_*/$M$_\odot$ is the mass and $a_* \equiv Jc/GM^2$ is
the spin on the neutron star. All current equations of state yield
$0.1 < a_* < 0.4$ (Friedman, Parker \& Ipser 1986; Cook, Shapiro \&
Teukolsky 1994), and we assume $m = 1.4$.

Since we do not observe the QPO in all of our observations, we use the empirical 
relation between the break and QPO frequencies from the fit to the data, 
\begin{equation} \nu_{\rm break} = 0.10 \, \nu_{QPO}^{1.10} {\rm Hz}, 
\end{equation} to estimate all the QPO frequencies. This, together with equation 
(\ref{eq:qpo}) implies the inner disc radius varying between 17--29$R_g$ 
($a_* = 0.1$) or 26--45$R_g$ ($a_* = 0.4$). These rather large radii are 
consistent with the rather small amount of relativistic smearing of the 
reflected spectral features which require that the disc is always larger than 
$20 R_g$ (see Section \ref{sec:phase-averaged}). 

\begin{figure}
\begin{center}
\leavevmode
\epsfxsize=8cm \epsfbox{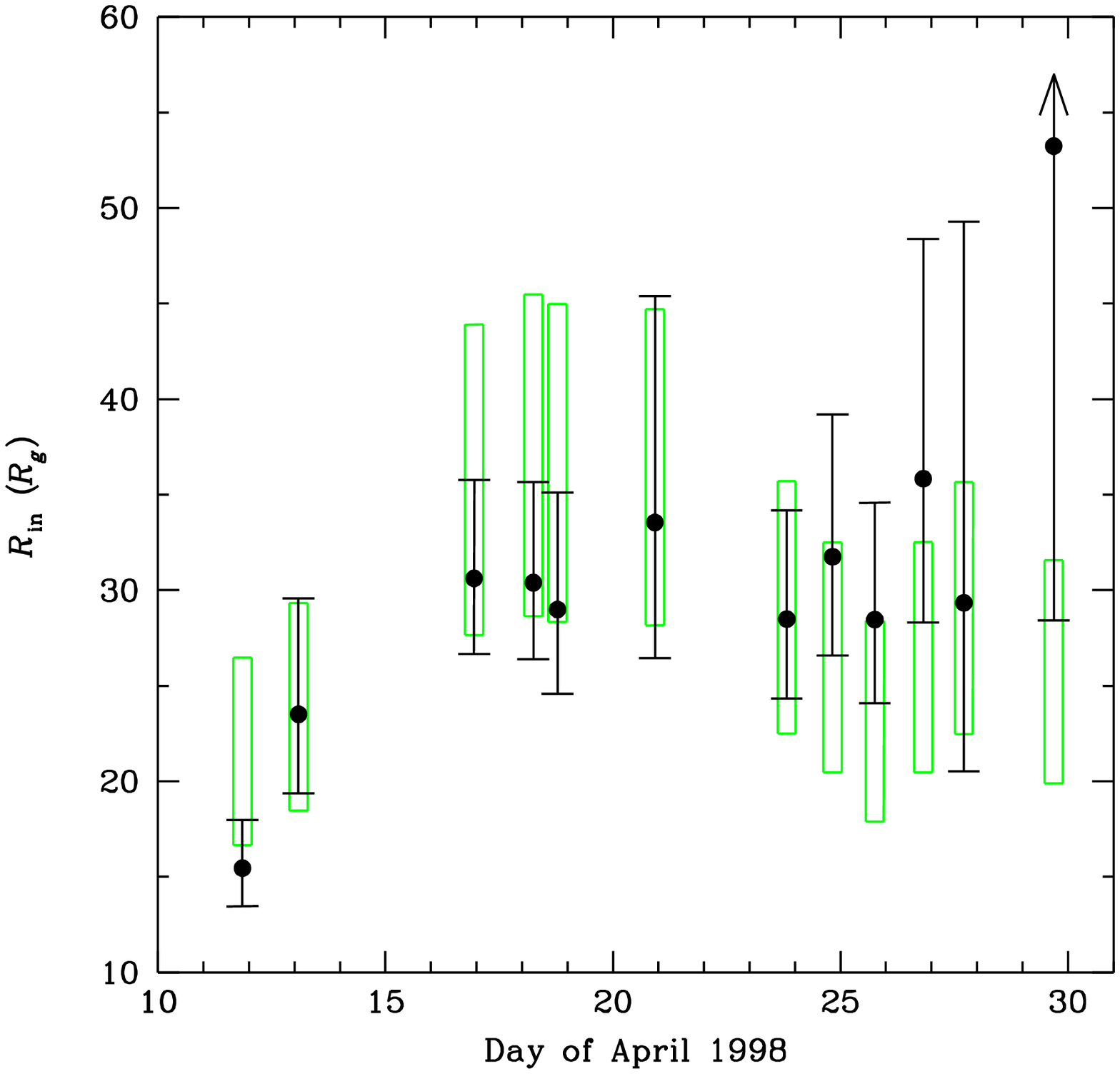}
\end{center}
\caption{The inner disc radius. Squares indicate the inner radius range
  for $0.1 < a_* < 0.4$, predicted from the fast variability (Equation
  \ref{eq:qpo}). Filled circles (with error bars) show the inner disc
  radius derived from the observed amplitude of reflection (see Fig.\
  \ref{fig:fit_params}), assuming that a half of the disc is
  irradiated by the shock just above the neutron star surface.}
\label{fig:radref}
\end{figure}

The solid angle subtended by the disc as measured by the reflected
spectrum gives us an independent way to estimate the inner disc
radius. The hard X-ray emission can only illuminate half the disc  in 
our  geometry where most of the hard X-ray emission comes
from a short shock above the magnetic pole. Then we derive the inner
disc radius as shown in Fig.\ \ref{fig:radref}, which is consistent
with the radii derived from the QPO frequencies.

The behaviour of the inner disc radius is rather complex. It evolves 
on the long time-scale of days. It first recedes as the outburst 
declines, then comes in again and then recedes (Fig.\ 
\ref{fig:radref}), rather than being uniquely related to the 
instantaneous mass accretion rate onto the compact object.  This lack 
of correlation with the mass accretion rate is generally seen in 
other LMXBs, e.g.\ 4U 1608--52 (M{\'e}ndez et al.\ 1999), showing 
that the key determinant of the behaviour of the system is tied 
instead to something with a longer time-scale, most probably the 
overall evolution of the disc.

\label{lastpage}

\end{document}